\documentclass[twocolumn,showpacs,preprintnumbers,amsmath,amssymb]{revtex4-1}

\usepackage{epsfig}
\usepackage{dcolumn}
\usepackage{bm}
\usepackage{xcolor}
\usepackage{amsmath}
\usepackage{float}
\usepackage{siunitx}
\usepackage{tikz}

\usepackage[normalem]{ulem}

\begin{document}

\preprint{\today} 

\title{Electronic phase-locking for three-color, two-pathway coherent control}

\author{Jonah A. Quirk$^{1,2}$,  Carol E. Tanner$^3$ and D. S. Elliott$^{1,2,4}$}

\affiliation{%
   $^1$Department of Physics and Astronomy, Purdue University, West Lafayette, Indiana 47907, USA\\
   $^2$Purdue Quantum Science and Engineering Institute, Purdue University, West Lafayette, Indiana 47907, USA\\
   $^3$Department of Physics and Astronomy, University of Notre Dame, Notre Dame, Indiana 46556, USA\\
    $^4$The Elmore Family School of Electrical and Computer Engineering, Purdue University, West Lafayette, Indiana 47907, USA
   }

\date{\today}

\begin{abstract}
We report a new method of two-pathway coherent control using three narrow-band cw laser sources, phase locked in an optical phase-lock loop, to maintain the high degree of optical coherence required for the coherent control process. In addition, we derive expressions for two-photon transition amplitudes and demonstrate their dependence on the polarization of the field components. This phase-locking technique expands the set of interactions to which coherent control techniques may be applied. It also allows for a constant low-frequency offset between the optical interactions, producing a continuous and constant phase ramp between the interactions, facilitating phase-sensitive detection of the modulating atomic signal. We illustrate this technique with two-photon vs.~one-photon excitation of a $\Delta F = 1$ component, and alternatively a $\Delta F = 0$ component, of the $6s \: ^2S_{1/2} \rightarrow 7s \: ^2S_{1/2}$ transition of atomic cesium.
\end{abstract}

\maketitle 

\section{Introduction}\label{sec:introduction}

In two-pathway coherent control, an atomic or molecular transition from a single initial state to a common final state is excited via two distinct coherent optical interactions, and the net outcome of the excitation can be controlled by varying the relative phase between the optical fields that drive the transitions.  Coherent control has been used to control the multiphoton ionization rate in atomic mercury~\cite{chen1990interference}, small molecules HCl and CO~\cite{lu1992coherent} and H$_2$S~\cite{kleiman1995coherenta}, and a larger molecule CH$_3$I~\cite{xing1996modulation}; measure the Guoy phase of a focused Gaussian beam as it propagates through the focal region~\cite{chen1990measurements}; control the angular distribution of photoelectrons in one- versus two-photon ionization of atomic sodium~\cite{PismaZhETF.55.431} and rubidium~\cite{yin1992asymmetric,wang2001determination} and molecular NO~\cite{yin1995two}; control the dissociation vs.~ionization rate in molecular HI~\cite{zhu1995coherent}; and control the phase lag between photoionization channels~\cite{gordon1999coherent,yamazaki2007observation,wang2021resonant}.  Recently, we have been pursuing precision measurements of weak optical interactions using an atomic homodyne detection technique based upon two-pathway coherent control~\cite{gunawardena2007weak,gunawardena2007atomic,antypasm12013,antypase2014,damitz2024technique,quirk2024measurement}.  In each of these examples, the coherence between the different optical pathways, which is a necessary condition for interference, was ensured by generation of the higher-frequency optical field using nonlinear frequency conversion (second-harmonic generation (SHG) or third-harmonic generation (THG)), for either pulsed or cw applications. Nonlinear optical conversion is an inherently coherent process, assuring a fixed relative phase difference between the harmonic and fundamental fields.

Recent ($\omega$, $2 \omega$) coherent control measurements in our laboratory~\cite{antypase2014, antypasm12013,quirk2024measurement} have been carried out using concurrent two-photon (frequency $\omega$) and one photon (frequency $2\omega$) excitation of the $6s \rightarrow 7s$ transition in cesium. (We use the abbreviated notation $ns$ for the $ns \: ^2S_{1/2}$ state.  Similarly, we will abbreviate the $6p \: ^2P_J$ level as $6p_J$.)  The interference between the amplitudes for these two processes is the basis for the coherent control measurement.  The light at frequency $\omega$ is a high-power infra-red cw beam at 1079 nm, produced by a low-power external cavity diode laser (ECDL) and a rare-earth-doped fiber amplifier.   
This configuration is capable of generating $>$10 W of narrow-band light at 1079 nm. Selection rules for one-color, two-photon excitation of an $ns \rightarrow n's$ transition, however, restrict the excitation to $\Delta F = 0$, $\Delta m =0$ transitions ~\cite{bonin1984two,marx1978wavelength, melikechi1986two}. ($\mathbf{F} = \mathbf{I} + \mathbf{J}$ is the total angular momentum of the atom, including nuclear spin $\mathbf{I}$ and total electronic angular momentum $\mathbf{J}$.  $m$ is the projection quantum number.)  
Since interference between transitions occurs only when the initial and final states of the atom are common to both optical pathways, only weak interactions on $\Delta F = 0$, $\Delta m =0$ transitions could be investigated.  That is, our previous technique was limited by its inability to measure these weak interactions on hyperfine-state-changing (i.e. $\Delta F = \pm 1$) transitions.
For this reason, we developed the technique described below to generate a two-color, two-photon excitation of the $6s \rightarrow 7s$ transition, and phase lock the green light (i.e., the $2 \omega$ beam at a wavelength of 540 nm) needed to drive the weak excitation and Stark-induced excitation pathways. 

In this paper, we report our technique for generating the optical fields necessary for the three-color two-pathway coherent control process; that is, electronic phase locking of three otherwise independent cw laser sources.  By this means, we are afforded greater flexibility in the choice of wavelengths and/or polarizations of the fields driving the optical interactions, allowing an expanded set of coherent control applications.  In the following section, we derive an expression for the two-photon transition amplitude for the $ns \rightarrow n's$ transition, which presents terms applicable when the polarization of the two fundamental field components are parallel to one another or, conversely, perpendicular to one another.  In the next section, we describe the technique for generating and phase-locking the optical fields, and in Section IV, we show results for the interference between a Stark-induced single-photon transition and a two-photon transition in cesium that requires laser fields that are polarized orthogonal, or alternatively parallel, to one another. 

\section{Two-photon absorption with parallel or perpendicular linear polarization components}\label{sec:tpselectionrules}

We stated earlier that $\Delta F = \pm 1$ components are forbidden for two-photon excitation of a $ns \: ^2S_{1/2} \rightarrow n's \: ^2S_{1/2}$ transition by a single-frequency field. In this section, we derive expressions for the amplitude of this two-photon transition. These expressions show this result explicitly.

\subsection{Two-Color Two-Photon Excitation}
Using time-dependent perturbation theory for interactions harmonic in time carried out to second order, we derive an expression for the two-photon excitation rate from an initial atomic ground state, $A$, through a virtual intermediate state, $n$, to a final excited state $B$ where we have also included the spontaneous emission decay rates, $\Gamma_n$ and $\Gamma_B$, of the intermediate and final states, respectively, as in the approach found in Ref.~\cite{Sakurai1967AdvQM}.

As a first step, we derive the second-order two-photon steady-state amplitude, $c_B^{(2)}$, of the excited state, $B$,
where the time of observation, $T$, is much greater than the lifetimes, $1/\Gamma_n$ and $1/\Gamma_B$, of the intermediate and excited states, respectively, such that the transient terms have decayed away,
\begin{eqnarray}\label{eq:c_B}
    c_B^{(2)} = \left(\frac{1}{i\hbar}\right)^2
    \frac{e^{i(\omega_B-\omega_A-\omega_2-\omega_1)T}}
    {i(\omega_B-\omega_A-\omega_2-\omega_1-i\Gamma_B/2)} \nonumber \\
    \times \sum_n \left( \frac{\langle B | V_2 | n \rangle \langle n | V_1 | A \rangle }{i(\omega_n - \omega_A  - \omega_1 -i\Gamma_n/2)} \right.\\
    \left. + \frac{\langle B | V_1 | n \rangle \langle n | V_2 | A \rangle }
    {i(\omega_n - \omega_A - \omega_2 -i\Gamma_n/2)} \right) . \nonumber 
\end{eqnarray}
Here $\omega_A = E_A/\hbar$, $\omega_B = E_B/\hbar$, and $\omega_n = E_n/\hbar$, where $E_A$, $E_B$, and $E_n$ are the energies of the initial state $A$, the final state $B$, and the intermediate states $n$, respectively. $V_i$ represents the interaction between one of the photon frequency components, $i=1$ or $2$, and the atom. The summation is taken over all states $n$ that are connected by $V_i$ to states $A$ and $B$. The two-photon amplitude can be enhanced by decreasing the detuning $|\omega_{np_J} - \omega_A - \omega_i| $ of one field component or the other from resonance with the intermediate state.  This enhancement can allow for strong two-photon absorption even with modest diode laser powers.  The amplitude in Eq.~(\ref{eq:c_B}) agrees with the matrix elements found in Ref.~\cite{Weissbluth1978Atoms}, although the spontaneous emission decay rates were not included in that treatment.  

In steady state equilibrium, the two-photon transition rate, $\mathcal{W}$, from A to B is equal to the decay rate of the excited state, B,
 \begin{equation}
     \mathcal{W}=  \Gamma_B |c_B^{(2)}|^2 .
 \end{equation}
Combining and rearranging, we have the two-photon transition rate from A to B,

\begin{eqnarray}
    \mathcal{W}  &=& 2\pi \left( \frac{1}{\hbar}\right)^4  \left| \sum_n \left( \frac{\langle B | V_2 | n \rangle \langle n | V_1 | A \rangle }{\omega_n - \omega_A  - \omega_1 -i\Gamma_n/2}  \right. \right. \nonumber \\
    &&  \hspace{0.3in}  \left. \left. + \frac{\langle B | V_1 | n \rangle \langle n | V_2 | A \rangle }{\omega_n - \omega_A  - \omega_2 -i\Gamma_n/2} \right) \right|^2   \\ 
    && \hspace{-0.1in}  \times \left\{ \frac{\Gamma_B/(2\pi) }{\left[ (\omega_B - \omega_A) - (\omega_1 + \omega_2)  \right]^2 + (\Gamma_B/2)^2} \right\}. \nonumber
\end{eqnarray}

In our case, the optical field consists of two frequency components, $\omega_1$ and $\omega_2$, such that the sum of these frequencies is near resonance with the atomic excitation $(\omega_1 + \omega_2) \approx (\omega_B - \omega_A)$. We write this as the familiar Fermi's Golden Rule
\begin{equation}
    \mathcal{W}  = \frac{2\pi}{\hbar}   \left| A_{2p} \right|^2  \tilde{\rho}_B (E)
\end{equation}
where
\begin{eqnarray}\label{eq:A2pasSumOverStates}
    A_{2p} &=&  \sum_n \left( \frac{\langle B | V_2 | n \rangle \langle n | V_1 | A \rangle }{E_n - E_A  - \hbar\omega_1 - i\hbar\Gamma_n/2}  \right.  \\ 
    & & \hspace{0.5in} \left. +\frac{\langle B | V_1 | n \rangle \langle n | V_2 | A \rangle }{E_n - E_A  - \hbar\omega_2 - i\hbar\Gamma_n/2} \right)  \nonumber
\end{eqnarray}
is the two-photon transition amplitude, and 
\begin{equation}
    \tilde{\rho}_B (E) = \frac{\hbar\Gamma_B/(2\pi) }{\left[ (E_B - E_A) - \hbar(\omega_1 + \omega_2)  \right]^2 + (\hbar\Gamma_B/2)^2} 
\end{equation}
is the density of states of the excited state.

Next, we write the interaction in terms of the length gauge 
\begin{equation}
    V_i = -q \mathbf{E}_i \cdot \mathbf{r}
    \:= e  E_i  \boldsymbol{\hat{\varepsilon}}_i \cdot \mathbf{r}
\end{equation} 
where $\mathbf{r}$ is the position operator of the electron, $q=-e$ is the charge of the electron, $e$ is the fundamental unit of charge, $\boldsymbol{\hat{\varepsilon}}_i$ is the photon polarization unit vector, and $E_i$ is the photon electric field amplitude. Some authors \cite{Grynberg2010QuantOpt} refer to this as the G{\"o}ppert-Mayer gauge \cite{goppert2009elementary}, and others \cite{reiss2014tunnelling} have noted that it does not satisfy the Lorenz gauge condition. Alternatively, the interaction can be written in the momentum gauge  
\begin{equation}
    V_i=-\frac{q}{m_e} \boldsymbol{\mathcal{A}}_i \cdot \mathbf{p}
    = \frac{e}{m_e} \mathcal{A}_i \boldsymbol{\hat{\varepsilon}}_i \cdot \mathbf{p},
\end{equation}
where $\mathbf{p}$ is the electron momentum operator, $m_e$ is the mass of the electron, and $\boldsymbol{\mathcal{A}}_i$ is the vector potential in the transverse gauge  ($\mathbf{\nabla} \cdot \boldsymbol{\mathcal{A}}=0$) and satisfies the Lorenz gauge condition \cite{Jackson1999,Jackson_2001}.  In the long-wavelength approximation where $e^{i\mathbf{k}_i\cdot \mathbf{r}} \approx 1$, the equivalence of the length and momentum gauges in two-photon absorption has been shown in Refs.~\cite{bassani1977choice,cohen1989photons}, who also discuss the more rapid convergence in the summation over intermediate states of the length gauge formalism.

\subsection{Evaluation of Angular Momentum Matrix Elements}

For the transition between two specific states specified by electronic angular momentum $J$, nuclear spin $I$, and total angular momentum $F$ with projection $m$ along the $\hat{z}$-axis, the electric dipole moment connecting two states (designated as primed and unprimed) can be reduced using~\cite{zare1988angular} 
\begin{eqnarray}\label{eq:rme1}
    \langle \gamma' J' I' F' m' | r | \gamma J I F m \rangle &=& (-1)^{F' - m'}  \\
    && \hspace{-1in}  \times \left( \begin{array}{lll} F' \quad 1 \quad F \\ m' \quad q_r \quad m \end{array}  \right)  \langle \gamma' J' I F' | r | \gamma J I F \rangle.\nonumber
\end{eqnarray}
$\gamma$ represents other quantum numbers of the states. The terms inside the parentheses are Wigner $3j$-symbols, which can be evaluated numerically using a number of applications or tables. $q_r$ indicates the component of the spherical irreducible tensor, with values $0$ or $\pm 1$.

We must also relate the components $\langle \gamma' J' I F' | r | \gamma J I F \rangle$ to the reduced matrix elements $\langle \gamma' J' || r || \gamma J \rangle$ using~\cite{zare1988angular} 
\begin{eqnarray}
    \langle \gamma' J' I F' | r | \gamma J I F \rangle &=& (-1)^{J' + I + F +1 } \nonumber  \\ 
    && \hspace{-0.5in} \times \sqrt{(2F' + 1)(2F + 1)}  \\ 
    && \hspace{-0.5in} \times \left\{ \begin{array}{lll}J' \quad F' \quad I \\ F \quad J \quad 1 \end{array}  \right\} \langle \gamma' J' || r || \gamma J \rangle. \nonumber    
\end{eqnarray}
The terms inside the curly brackets are Wigner $6j$-symbols, which can also be evaluated numerically through the same applications or tables.

For two-photon excitation when both frequency components are polarized parallel to one another, we define the $\hat{z}$ direction of the system along this polarization direction, so $\boldsymbol{\hat{\varepsilon}}_1 = \boldsymbol{\hat{\varepsilon}}_2 = \boldsymbol{\hat{z}}$.  Then 
\begin{equation}
    \boldsymbol{\hat{\varepsilon}}_i \cdot \mathbf{r}  =  \boldsymbol{\hat{z}} \cdot \mathbf{r} =  r_0
\end{equation}
for $i = 1$ and $2$, $z = r_0 $ when written in a spherical basis set, and 
$q_r$ is equal to zero for both applications of Eq.~(\ref{eq:rme1}). We define the $\boldsymbol{\hat{y}}$ direction as the propagation direction of both laser fields, $\boldsymbol{\hat{y}} = \hat{\mathbf{k}}_1 = \hat{\mathbf{k}}_2$.  (The $\omega_1$ and $\omega_2$ beams are co-propagating in our application, as is required  for phase matching with the one-photon interaction in the coherent control process.) In the limit when neither field frequency is close to resonance with the intermediate state, i.e. $|\omega_j - \omega_n|$ is much greater than $\Gamma_n$ as well as the frequency spacing between hyperfine components of the transition, we can ignore the energy difference between hyperfine components of the intermediate states. Then, evaluating the Wigner-3j and 6j symbols for the $6s \: ^2S_{1/2} \rightarrow 7s \: ^2S_{1/2}$ transition of cesium (with nuclear angular momentum $I=7/2$) through the virtual $np \: ^2P_{J}$ the $J = 1/2$ and $3/2$ intermediate fine structure states individually, the transition amplitude Eq.~(\ref{eq:A2pasSumOverStates}) for parallel polarizations ($\parallel$) can be written 
\begin{equation}\label{eq:A2pparpol}
    A_{2p}^{\parallel} = \tilde{\alpha} E_1 E_2 \delta_{F_A,F_B} \delta_{m_A,m_B},
\end{equation}
where
\begin{eqnarray}
    \tilde{\alpha} &=& \frac{1}{6} \left[ - \sum_n  \left(\frac{1}{\omega_{np_{1/2}}-\omega_1} + \frac{1} {\omega_{np_{1/2}}-\omega_2} \right) \right.  \\ 
    && \hspace{0.8in} \times 
    \langle 7s || r || np_{1/2} \rangle  \langle np_{1/2} || r || 6s \rangle \rule{0in}{0.25in} \nonumber \\
    &&  \hspace{0.2in} + \sum_n \left(\frac{1}{\omega_{np_{3/2}}-\omega_1} + \frac{1}{\omega_{np_{3/2}}-\omega_2} \right)  \nonumber  \\
    && \left. \hspace{0.8in} \times \langle 7s || r || np_{3/2} \rangle  \langle np_{3/2} || r || 6s \rangle  \rule{0in}{0.25in}\right]. \nonumber
\end{eqnarray}
Note that when the polarizations of both field components are parallel to one another, only $\Delta F = 0$, $\Delta m = 0$ transitions are allowed. This condition is always satisfied when the two-photon transition is driven by a single frequency field, i.e. $\omega_2 = \omega_1$. That is, one-color two-photon excitation on this transition cannot include $\Delta F = \pm 1$ or $\Delta m = \pm 1$ components.

In order to excite $\Delta F = \pm 1$ transitions, one can employ cross-polarized field components.  Choosing $\boldsymbol{\hat{z}}$ along one field component polarization ($\boldsymbol{\hat{\varepsilon}}_1 = \boldsymbol{\hat{z}}$), $\boldsymbol{\hat{x}}$ for the other ($\boldsymbol{\hat{\varepsilon}}_2 = \boldsymbol{\hat{x}}$), and both directions of propagation along $\boldsymbol{\hat{y}} = \hat{\mathbf{k}}_1 = \hat{\mathbf{k}}_2$, 
\begin{eqnarray}
\boldsymbol{\hat{\varepsilon}}_1 \cdot \mathbf{r} 
=\boldsymbol{\hat{z}}\cdot \mathbf{r} 
= r_0\\
\boldsymbol{\hat{\varepsilon}}_2\cdot\mathbf{r}
=\boldsymbol{\hat{x}}\cdot\mathbf{r}
= \frac{1}{\sqrt{2}}(r_{-1}-r_{+1}),
\end{eqnarray}
so we use $q_r = 0$ in evaluating the $3j$ symbols of Eq.~(\ref{eq:rme1}) for field 1, and $q_r = \pm1$ for field 2. 
The spherical basis position variables $r_{\pm1}$ are $r_{+1} = -(x+iy)/\sqrt{2}$ and $r_{-1} = (x-iy)/\sqrt{2}$. 
The two-photon amplitude for perpendicular polarizations ($\bot$) reduces to
\begin{equation}\label{eq:A2pperppol}
    A_{2p}^{\bot} = \tilde{\beta} E_1 E_2 C_{F_A,m_A}^{F_B,m_B},
\end{equation}
where
\begin{eqnarray}\label{eq:betatilde}
    \tilde{\beta} &=& \frac{1}{6} \left[ \sum_n \left(\frac{1}{\omega_{np_{1/2}}-\omega_1} - \frac{1} {\omega_{np_{1/2}}-\omega_2} \right) \right.  \\
    && \hspace{0.8in} \times 
    \langle 7s || r || np_{1/2} \rangle  \langle np_{1/2} || r || 6s \rangle \rule{0in}{0.25in} \nonumber \\
    &&  \hspace{0.1in} + \frac{1}{2} \sum_n \left(\frac{1}{\omega_{np_{3/2}}-\omega_1} - \frac{1}{\omega_{np_{3/2}}-\omega_2} \right)  \nonumber \\
    && \left. \hspace{0.8in} \times \langle 7s || r || np_{3/2} \rangle  \langle np_{3/2} || r || 6s \rangle  \rule{0in}{0.25in}\right]. \nonumber
\end{eqnarray}
The factors $C_{F_A,m_A}^{F_B,m_B}$, which are related to the Clebsch-Gordon coefficients, are derived from angular momentum algebra, and are identical to those given in Ref.~\cite{GilbertW1986}, which dealt with Stark-induced transitions on the same transition.  Note that inside the parentheses in either of the summations within Eq.~(\ref{eq:betatilde}), we find the difference between $(\omega_{np_J}-\omega_1)^{-1}$ and $(\omega_{np_J}-\omega_2)^{-1}$.  For $\omega_1 = \omega_2$, this term vanishes.  Thus when driving the two-photon interaction with a single frequency field with $\omega_1 = \omega_2 \approx (\omega_B - \omega_A)/2$, the only contribution to the two-photon excitation is given by Eq.~(\ref{eq:A2pparpol}), which permits only $\Delta F = 0$, $\Delta m = 0$ transitions.  From this we conclude that in order to drive $\Delta F = \pm 1$ components of this two-photon transition, it is necessary to use non-degenerate laser frequencies $\omega_2 \neq \omega_1$, motivating the three-color coherent control technique described in this paper.  

We summarize the total two-photon transition amplitude to be,
\begin{align}\label{eq:A2total_alt}
    A_{2p} &= \tilde{\alpha} \mathbf{E}_1 \cdot \mathbf{E}_2  \: \delta_{F_A,F_B} \delta_{m_A,m_B}\nonumber\\
    &\hspace{1cm}+\tilde{\beta} \left( \mathbf{E}_1 \times \mathbf{E}_2 \right) \cdot \hat{y} \:  C_{F_A,m_A}^{F_B,m_B},
\end{align} 
presuming that all lasers propagate in the same direction, as is necessary for uniform interference throughout the interaction region. There are interesting parallels between this expression for the two-photon absorption amplitude and that for Stark-induced transitions~\cite{bouchiat1975parity,GilbertW1986}.

\section{Two-Pathway Coherent Control}\label{sec:TPCC}
\begin{figure}[b!]
\begin{center}
    \includegraphics[width=0.45\textwidth]{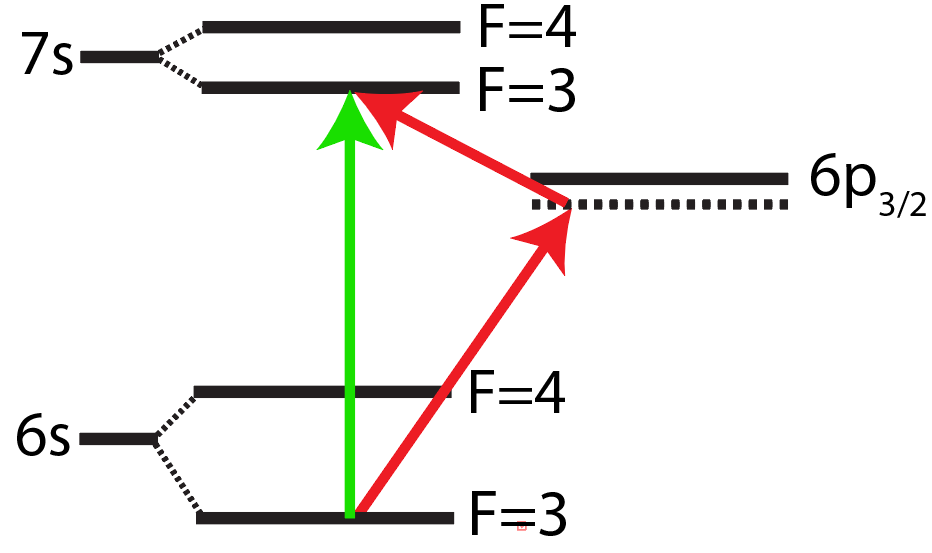}
\caption{Energy level diagram of the relevant energy levels of cesium in the coherent control interaction.  The $6s \rightarrow 7s$ transition is driven via a single-photon ($\lambda = 539.5$ nm) interaction, represented by the green arrow, and concurrently by the two-photon interaction, represented by the red arrows (one photon at 852 nm, which is close to, but not resonant with, the $6s \rightarrow 6p_{3/2} $ transition; the second at 1470 nm).  The transition shown in this plot is an example of a $\Delta F = 0$ transition. }
\label{fig:energyleveldiagram}
\end{center}
\end{figure}

In our current applications~\cite{antypasm12013,antypase2014,quirk2024measurement}, we use coherent control to measure the transition moments of extremely weak single photon transitions between the $6s$ and the $7s$ states of atomic cesium. See Fig.~\ref{fig:energyleveldiagram} 
for an energy level diagram of cesium showing the relevant states and transitions.
Among the weak transitions under investigation in our laboratory are, for example, the magnetic dipole (M$_1$) transition and the weak-force induced electric dipole (E$_1$) transition.  
These transitions are known to consist of nuclear-spin-independent (NSI) as well as nuclear-spin-dependent (NSD) contributions~\cite{BennettRW99, dzuba2000off,wood1997measurement,haxton2001atomic}.  The latter couple states of different total angular momentum $F$ and $F'$.

Our measurement technique is based on the interference between one of these weak transitions, a controllable Stark-induced interaction~\cite{bouchiat1975parity, GilbertW1986}, and a strong two-photon interaction.  When each of these interactions couples the same initial and final states (including magnetic projection quantum number $m$), the net excitation rate of the excited $7s$ state can be written as
\begin{displaymath}
  \mathcal{W} = \frac{2 \pi}{\hbar} \left|\rule{0in}{0.15in} A_{\rm 2p} + A_{\rm St} + A_{\rm weak} \right|^2 \tilde{\rho}_{7s}(E),
\end{displaymath}
where $A_{\rm 2p}$, $A_{\rm St}$, and $A_{\rm weak}$ are the transition amplitudes for the two-photon, Stark, and weak interactions, respectively, and $\tilde{\rho}_{7s}(E)$ is the density of states of the $7s$ state.

In this work, in which we are focused only on a new experimental scheme for generating interference between a single-photon amplitude and a two-photon amplitude with electronically-phase-locked lasers, we will omit any further discussion of the weak interaction, and focus our attention to the interference between the two-photon amplitude $A_{\rm 2p}$ and the Stark-induced amplitude $A_{\rm St}$.

When the laser is resonant with the transition, the linewidth is lifetime limited, and the weak amplitude is negligible, the transition rate simplifies to
\begin{displaymath}
  \mathcal{W} = \frac{4}{\hbar^2 \Gamma} \left|\rule{0in}{0.15in} A_{\rm 2p} + A_{\rm St}  \right|^2 ,
\end{displaymath}
where $\Gamma$ is the decay rate of the $7s$ state.

In our typical application~\cite{antypasm12013,antypase2014,quirk2024measurement}, the Stark amplitude is much weaker than the two-photon amplitude $A_{\rm 2p}$, since we use the latter as the local oscillator in the `heterodyne' detection process and the former for calibration of the weak interaction amplitude.  Under this condition, the transition rate reduces to 
\begin{equation}\label{eq:Wcosphi}
  \mathcal{W} = \frac{4}{\hbar^2 \Gamma} \left[ \rule{0in}{0.15in} |A_{\rm 2p}|^2 + |A_{\rm 2p}||A_{\rm St} | \cos(\Delta \phi + \delta) \right],
\end{equation}
where we have omitted the term that is second order in $|A_{\rm St}|$. $\Delta \phi$ is the difference between the optical phase of the green beam and the sum of the phases of the two-photon fields, and $\delta$ is the constant phase due to the Stark amplitude $A_{\rm St}$. ($\delta = 0$ or $\pm\pi/2$, depending on the field and polarization configuration used in the measurement~\cite{bouchiat1975parity,GilbertW1986}.)

\begin{figure}[t!]
 \includegraphics[width=0.47\textwidth]{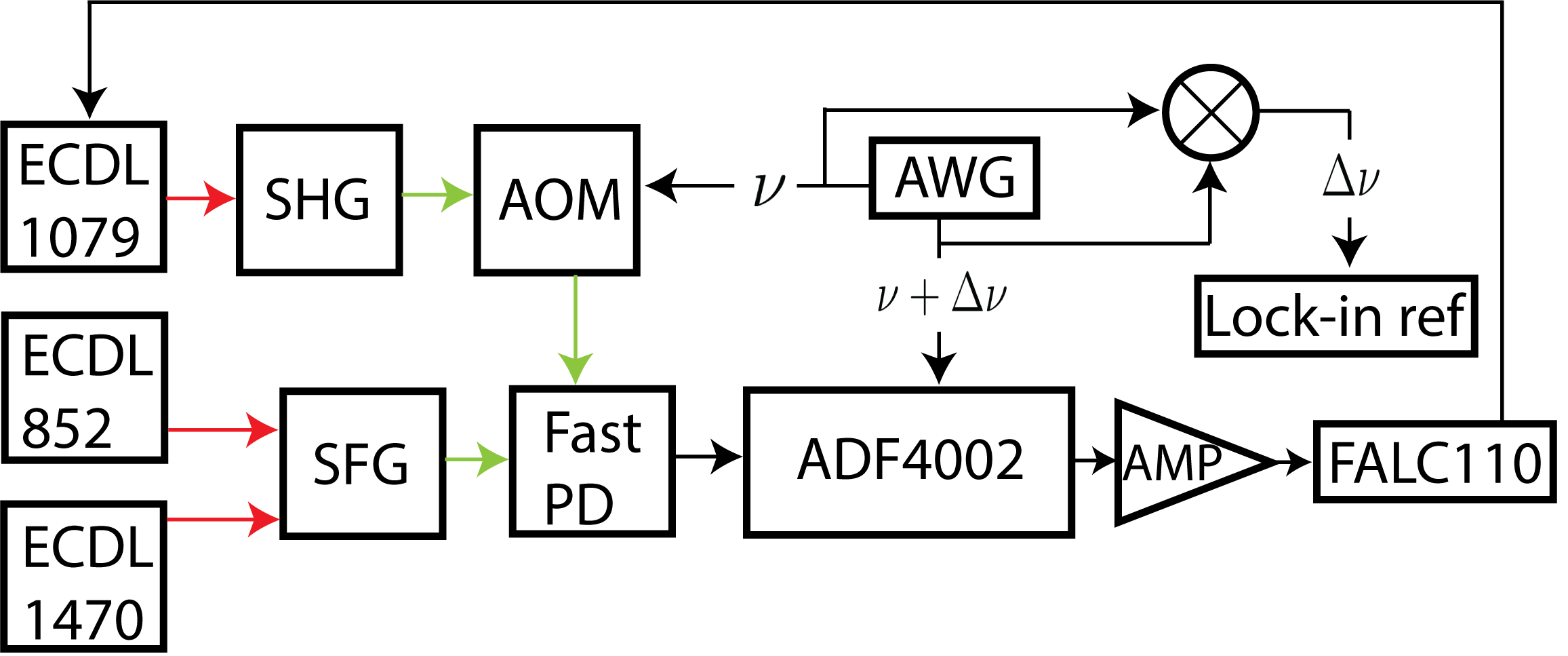}
 \caption{A high-level diagram depicting the technique that generates laser fields necessary to drive coherent one- and two-photon transitions. ECDL - external cavity diode laser, with the wavelength of the output (in nm) indicated for each, AOM - acousto-optic modulator, SHG - second harmonic generation crystal, SFG - sum-frequency generation crystal, AWG - arbitrary waveform generator, PD - photodiode, $\bigotimes$ - frequency mixer. The FALC110 is a fast servo and the ADF4002 chip is used to generate an error signal from the beatnote between the SHG beam and the SFG beam. We list here these specific instruments, not necessarily as an endorsement/requirement, but what we chose to use. A sufficiently fast servo or a comparable digital phase lock loop (PLL) chip would also work.}
 	  \label{fig:Phasecoherence}
\end{figure}

We show a schematic layout of the generation and phase-locking technique in Fig.~\ref{fig:Phasecoherence}.  
As described above, the primary laser source is a 1079 nm ECDL, whose output is amplified in a fiber amplifier.  The 1079 nm light is frequency doubled in a periodically-poled lithium niobate crystal (SHG).  We use this beam to drive the weak and Stark-induced linear interactions in the atomic beam, which require high optical power because their moments are so weak.  In the following, we refer to this beam as the SHG beam.

The beams that drive the two-photon interaction in the atom beam are generated by two ECDLs, one producing an output beam at a wavelength of 852 nm, the other at 1470 nm.  (See Fig.~\ref{fig:energyleveldiagram}.)  We stabilize the frequency of the 852 nm beam $\sim$ 1 GHz below the $F=2$ or 3 line of the $6s \rightarrow 6p_{3/2}$ transition using a saturated absorption resonance in a cesium vapor cell. The 1470 nm laser frequency is tuned to complete the two-photon transition to the excited $7s$ state.  We choose the 1 GHz detuning of the intermediate step of the two-photon resonance from the $6p_{3/2}$ line
to be small enough to enhance the two-photon interaction, but large enough to minimize the direct excitation of population of the intermediate $6p_{3/2}$ state, which would reduce the coherence of the two-photon excitation.

The 1470 nm laser frequency is stabilized to a frequency comb laser source (FCL) via an optical-phase lock loop (OPLL).  The carrier envelope offset frequency and repetition rate of the FCL are stabilized to a GPS-conditioned reference. The near-IR output of the FCL is spectrally filtered using a ruled reflective diffraction grating and coupled into an optical fiber. This filtered FCL output is combined with 3 mW of the 1470 nm laser beam on a 95:5 (FCL:1470 nm laser) fiber beam combiner. The combined FCL and laser light are beat on a 5 GHz InGaAs photodiode. This beat signal is amplified to greater than -10 dBm and injected into a homemade OPLL.  This loop consists of an ADF4002-evaluation board, loop filter, and 50 $\Omega$ line driver. The ADF4002 evaluation board acts as a divider, phase detector, and charge pump for phase discrimination. The output error signal of this OPLL is supplied to a fast servo to correct the 1470 nm laser frequency through current and piezo modulation. A reference frequency is supplied to the ADF4002-evaluation board to vary the offset lock point between the FCL and 1470 nm laser source. Once the 852 nm laser is stabilized 1 GHz below the $6s \rightarrow 6p_{3/2}$ transition, the 1470 nm laser is then locked offset from the comb such that the 852 nm and 1470 nm beams are resonant with the $6s \rightarrow 7s$ transition.

\begin{figure}[b!]
 \includegraphics[width=0.44\textwidth]{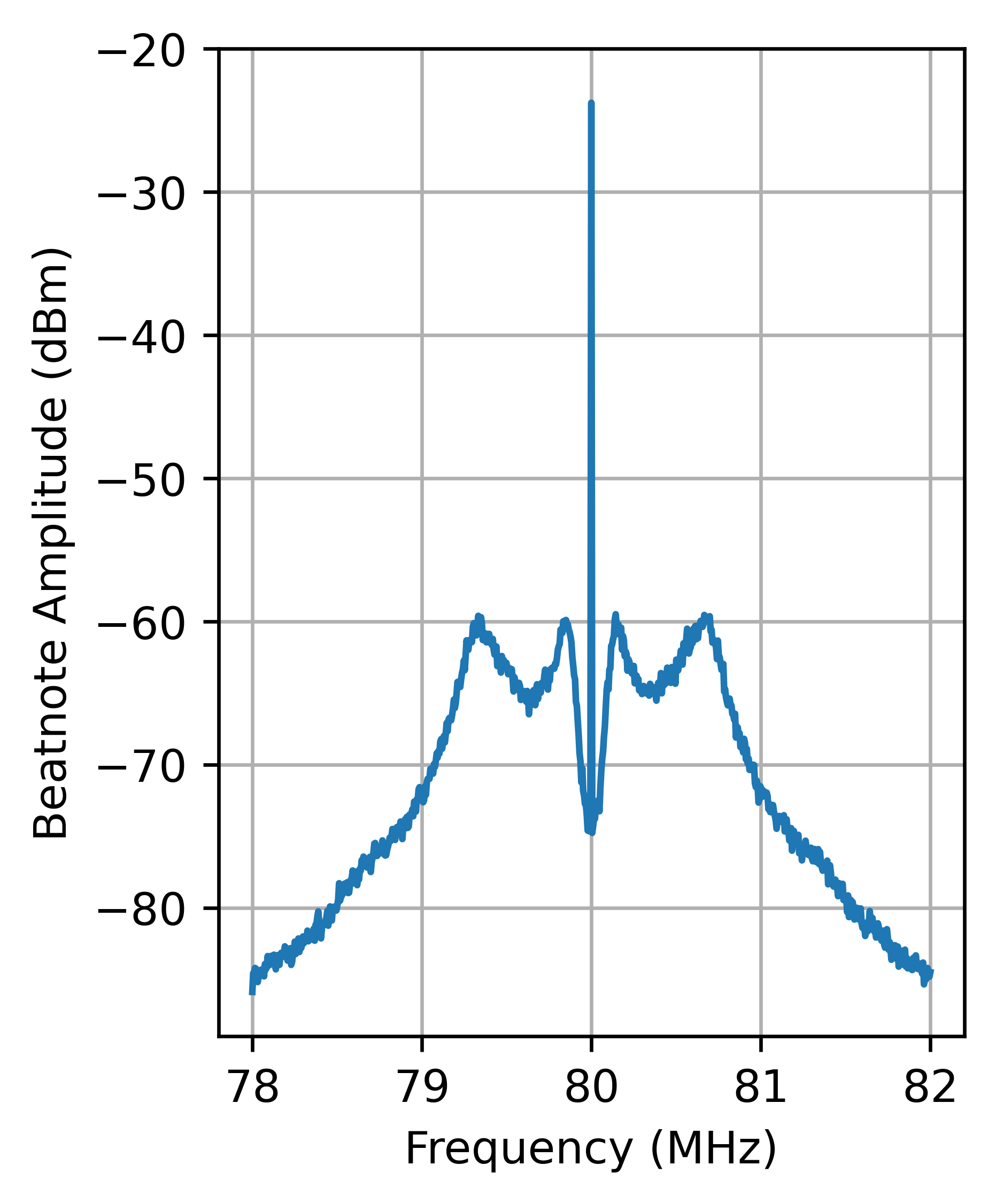}
 \vspace{-1\baselineskip}
 \caption{Phase-locked beatnote spectrum between the SHG and the SFG beams used to generate coherence between the two-photon and single-photon transitions. The resolution bandwidth is 1 kHz. }
 	  \label{fig:beatnote}
\end{figure}

To phase lock the two-photon interaction to the direct one-photon interaction, we combine the 852 nm and 1470 nm beams on a dichroic beamsplitter, and focus them into a 40 mm long magnesium-doped periodically-poled lithium niobate nonlinear crystal, generating a visible 540 nm beam using sum frequency generation (SFG).  With 120 mW of 852 nm light and 40 mW of 1470 nm light, we generate 0.2 mW of 540 nm light.  (We call this beam the SFG beam in the following.) We use the beat signal between the SFG beam and a beam derived from the SHG beam to stabilize the frequency of the 1079 nm source. This SHG component beam is produced by directing the primary SHG beam (1100 mW) through an acousto-optic modulator (AOM) that is driven at frequency $\nu = 80$ MHz (input rf power = 7 dBm) to generate a $\sim$11 mW diffracted beam frequency-shifted by $\nu$.  The AOM drive signal is generated by an arbitrary waveform generator (AWG).  We combine the SFG and frequency-shifted SHG beams on a 90:10 beam combiner and launch both beams into a fiber-coupled fast photodiode to observe the beat signal. When the frequencies of the SFG and unshifted SHG beams match one another, the beat signal frequency is 80 MHz. This 80 MHz beat signal is fed into an OPLL (similar to the one used to stabilize the 1470 nm laser) to generate an error signal that is used to phase-lock the SHG beam to the SFG beam. The power spectrum of this stabilized beat signal is shown in Fig.~\ref{fig:beatnote}. 
The reference input to the ADF4002-evaluation board chip is at a frequency $\nu - \Delta \nu$, where $\Delta \nu = 150 $ Hz.  The ADF4002-evaluation board and loop filter generates a phase error signal, which we feed back to the controller for the 1079 nm laser. This process allows us to phase stabilize the SHG beam to the two-photon beams, offset by 150 Hz. We chose 150 Hz for the offset frequency due to a longitudinal velocity spread in our atomic beam and the physical distance from the interaction region to the detection region in our system, where higher frequencies would wash out the interference. For other experimental configurations, this may not be a limiting factor and one could choose a different frequency.

We conduct this experiment in a high vacuum chamber containing an atomic beam and electric field assembly. Cesium atoms are prepared into the $6s, F=3$ ground state and then excited to the $7s, F=4$ or ($F=3$ for a $\Delta F=0$ transition) by the three phase-locked beams described above. Once excited, the atoms return to either hyperfine ground state. Atoms in the previously emptied ($F=4$) state are excited by the output of another 852 nm diode laser (called the detection laser) in a cycling transition where several fluoresced photons are collected on a large area photodiode. A transimpedance amplifier of gain \SI{20}{\Mohm} converts this photocurrent into a measurable voltage for a digital storage oscilloscope or a lock-in amplifier. A more detailed discussion of the atomic beam apparatus is included in \cite{quirk2024measurement}.

\begin{figure}[b!]
 \includegraphics[width=0.5\textwidth]{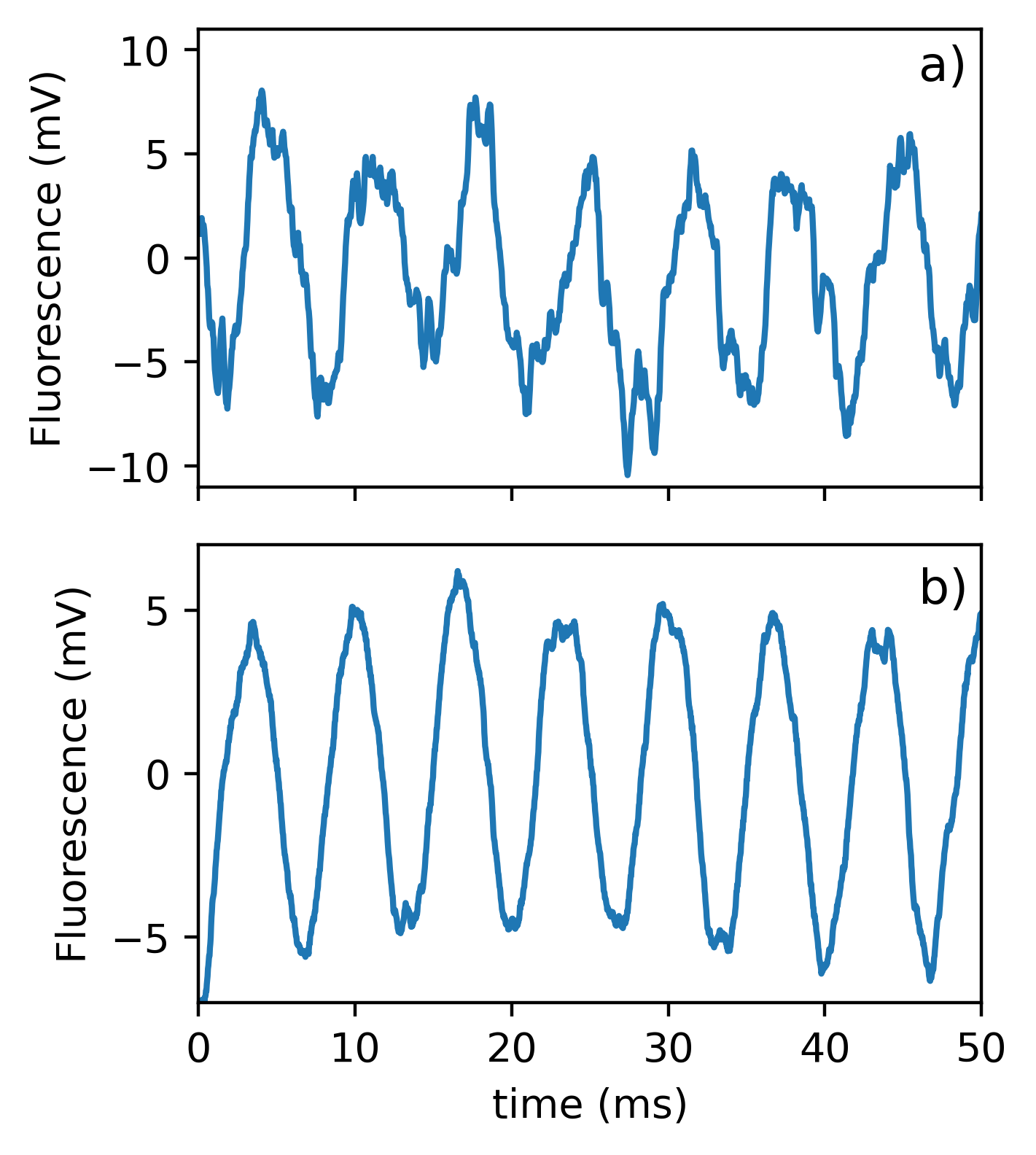}
 \vspace{-1\baselineskip}
 \caption{Representative example of the 150 Hz modulation of the atomic excitation signal produced when the interfering excitations (single-photon and two-photon) differ by a constant frequency offset of 150 Hz. Figure a) shows the interference directly in a single trace observed on a digital oscilloscope. Figure b) shows the same interference after averaging eight traces. Here the $\Delta \nu = 150 $ Hz reference signal is used as the trigger input to the oscilloscope.}
 	  \label{fig:atomicsignal150Hz}
\end{figure}

The 852 nm (5.2 mW) and 1470 nm (2.2 mW) beams are then carefully overlapped with the SHG beam (1100 mW) and all three beams are weakly focused onto the atomic beam (852 nm - 850 $\mu$m waist diameter, 1470 nm - 860 $\mu$m waist diameter, SHG - 560 $\mu$m waist diameter). A pair of electric field plates are centered on the three focused laser beams and atomic beam. These plates are oriented such that they produce an electric field along the laser propagation axis, $\hat{k}$. A small hole drilled in the center of each plate allows transmission of the laser beams.  A \SI{1}{\kV/\cm} field is applied to induce the one-photon Stark transition. When the optical beams are phase-locked at a constant 150 Hz offset, the interference between the two-photon interaction and the weak linear interactions is constantly varying at 150 Hz.  We show an example of the modulating interference signal in Fig.~\ref{fig:atomicsignal150Hz}.
Precise measurement of the amplitude of this oscillating atomic excitation signal, which results from the interference term introduced in Eq.~(\ref{eq:Wcosphi}), is the goal of these studies.  We measure this amplitude directly using a lock-in amplifier, where the reference signal to the lock-in amplifier is at the frequency $\Delta \nu = 150$ Hz.  This reference signal is generated with an rf mixer, whose inputs are the AOM drive signal at frequency $\nu$ and the reference input to the ADF4002 evaluation board at frequency $\nu - \Delta \nu$.   

Note the important additional benefit of the phase-locking technique: we eliminate the need for modulating or ramping the phase shift between the beams in an external galvo-mounted window (as we used previously~\cite{antypase2014, antypasm12013}).  The former technique required extremely linear phase ramps and nonlinear fits of sinusoidal functions to the measured data sets to determine the amplitude.  With the present technique, the output of the lock-in amplifier directly yields the amplitude of the interference signal, greatly simplifying the analysis.

\section{Results}\label{sec:Results}
Earlier, we stated that this three-color, two-pathway coherent control technique opens the possibility for observing and measuring interference on $\Delta F = \pm 1$ components, as well as $\Delta F = 0$ components, of the $6s \rightarrow 7s$ transition.  In this section, we demonstrate this capability.

\begin{figure}[t!]
 \includegraphics[width=0.43\textwidth]{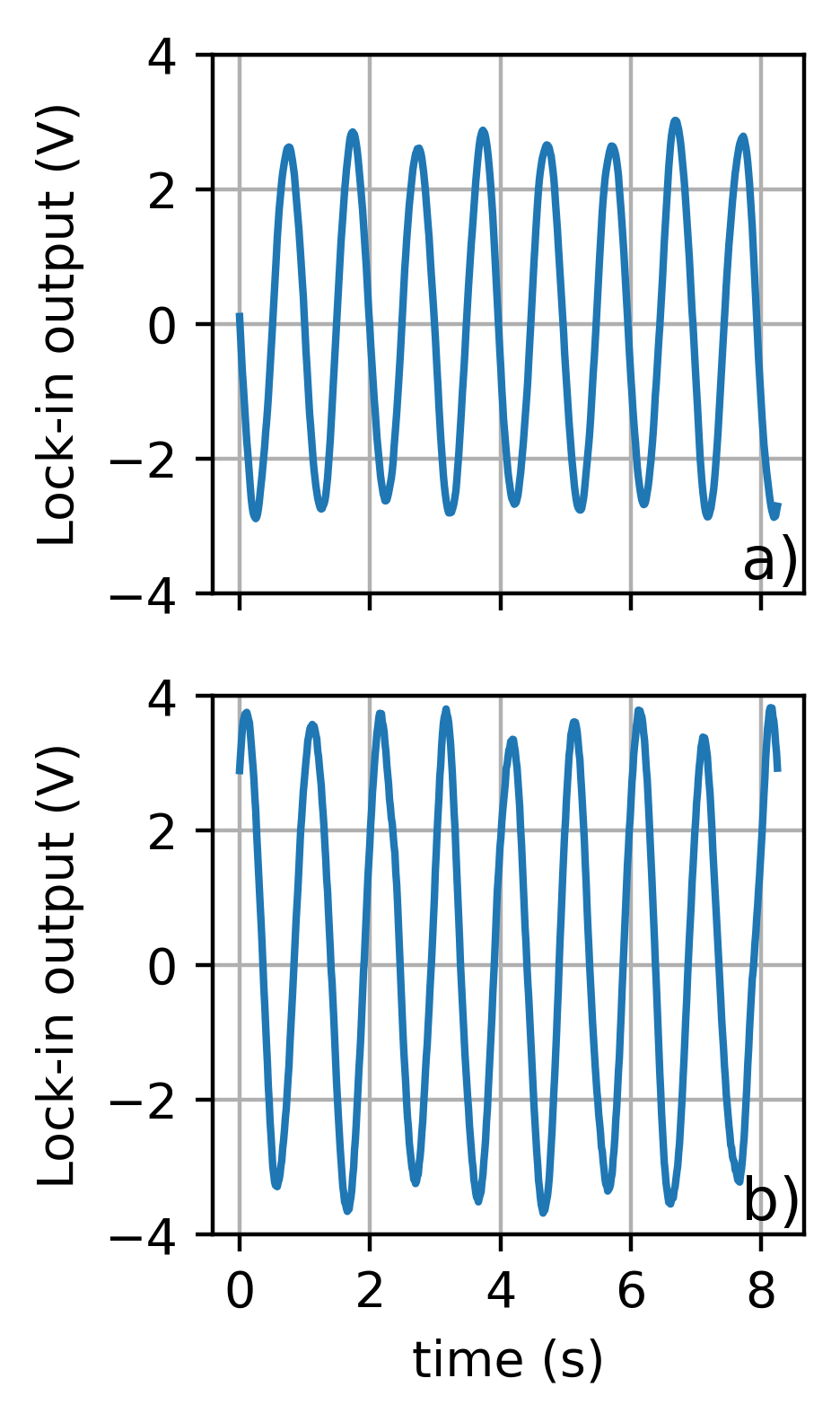}
 \vspace{-1\baselineskip}
 \caption{Representative examples of two-pathway coherent control signals collected while using the electronic phase-locking technique described in this report. Each of these plots illustrate interference between a Stark-induced transition (one-photon) and a two-photon transition.  a) A $\Delta F = 0$ transition ($F = 3 \rightarrow F^{\prime} = 3$), and b) a $\Delta F = +1$ transition ($F = 3 \rightarrow F^{\prime} = 4$).  In each case, we show the output of the lock-in amplifier as we ramp the phase difference between the single-photon and two-photon transitions at 1 Hz.}
 	  \label{fig:lockinoutput}
\end{figure}

In this experiment, we observe the interference between a Stark-induced one-photon transition and a two-photon transition on a $\Delta F =1$ line ($6s, F=3 \rightarrow 7s, F=4$). We show the amplitude of this interference signal, as measured with a lock-in amplifier, in Fig.~\ref{fig:lockinoutput}(b). The linear polarizations of the $\omega_1$ and $\omega_2$ beams are perpendicular to one another for this measurement. Since $\mathbf{E_1} \cdot \mathbf{E_2} = 0$, the $A_{2p}^{\perp}$ term of Eq.~(\ref{eq:A2pperppol}), or the second term in Eq.~(\ref{eq:A2total_alt}) determines the two-photon amplitude in this case.  This demonstration illustrates the new capability made possible through the three-color technique, in that two-photon absorption on the $\Delta F =1$ line of this transition vanishes when $\omega_1 = \omega_2$, as shown in Sec.~\ref{sec:tpselectionrules}. 

We show similar measurements, also using the three-color technique, on a $\Delta F =0$, $\Delta m =0$ line in Fig.~\ref{fig:lockinoutput}(a).  The traditional coherent control technique (i.e.~with $\omega_1 = \omega_2$) was applicable on $\Delta F =0$ transitions, of course, but it is convenient that the three-color technique introduced in this work is applicable for either $\Delta F=0$ or $\pm1$ transitions.  The only changes to the set-up needed are to tune the frequencies of the 1470  nm and 540 nm beams to the $6s, F=3 \rightarrow 7s, F=3$ transition frequency, and to align the polarizations parallel to one another.  With $\mathbf{E_1} \times \mathbf{E_2} =0$, the two-photon amplitude is determined by Eq.~(\ref{eq:A2pparpol}), or the first term in Eq.~(\ref{eq:A2total_alt}). The laser powers in the interaction region differ between Fig.~\ref{fig:lockinoutput}(a) and (b), so we should not assign any significance to the different amplitudes of these two signals, other than to note that they are comparable to one another.  Both traces show good phase control and low noise.

Interference between the Stark-induced and two-photon interactions relies on strict relative phase control of the excitation lasers. By scanning the relative phase of the \SI{540}{\nm} beam relative to the sum of the \SI{1470}{\nm} and \SI{850}{\nm} beams, the relative phase of the Stark induced transition and two-photon transition amplitudes is shifted as well, resulting in interference. Fig.~\ref{fig:atomicsignal150Hz} shows a direct measurement of the 150 Hz modulation in the collected fluorescence without the use of a lock-in amplifier. In Figs.~\ref{fig:atomicsignal150Hz}(a) and (b), we show the interference signal (a) with no averaging on an oscilloscope, and (b) when averaged over eight traces, for the $\Delta F = 0$ components. Fig.~\ref{fig:lockinoutput} shows the output of a lock-in amplifier using a 151 Hz reference to demodulate the signal down to near DC, allowing us to optimize the interference signal and demonstrate coherence. Figs.~\ref{fig:lockinoutput}(a) and (b) illustrate the demodulated signals at 1 Hz for $\Delta F = 0$ (a) and $\Delta F = \pm 1$ (b) components. Ultimately, a weak signal measurement would then consist of demodulating the interference signal directly down to DC, where the lock-in output would then be proportional to the interference amplitude.

We will describe precision measurements of the amplitude ratio $M_1/\beta$ in a future report.  Here the primary purpose is to show the strong interference signal using the three independent laser sources phase-locked to permit the interference.

\section{Conclusion}\label{sec:Conclusion} 
Using three-color two-pathway coherent control, the technique described in this paper, we have demonstrated the ability to interfere optical transitions and to vary the total transition rate by ramping the phase difference between the optical interactions.  While our direct motivation for developing this phase locking technique was for application to our precise measurements of very weak transition moments, including atomic parity violation measurements, it can also find application in excitation and control of atomic Rydberg states for potential use in efficient conversion of microwave to optical photons or in sensitive electric field sensing.

Here we choose 150 Hz modulation as it is a limitation of the atomic beam apparatus. Higher bandwidth modulation and detection could be realized in other experimental apparatus with a potential upper limit near the atomic linewidth. 

\section{Data Availability Statement}
The data that support the findings of this article are openly available~\cite{datafiles}.

\section{Acknowledgment}\label{sec:Acknowledgment}

This material is based upon work supported by the National Science Foundation under Grant Number 2409461-PHY.

\bibliography{biblio}

\end{document}